\documentclass[aps, prx, reprint, showpacs, superscriptaddress, longbibliography]{revtex4-1}
\usepackage{amsmath, amssymb, graphicx, subfigure, CJK} 
\usepackage[normalem]{ulem}
\usepackage{mathptmx, color, hyperref}
\usepackage[all]{hypcap}

\newcommand{\figurewidth}{0.5\textwidth}

\definecolor{MyDarkGreen}{rgb}{0,0.6,0}
\definecolor{MyDarkBlue}{rgb}{0,0,0.8}
\definecolor{MyDarkRed}{rgb}{0.6,0,0.3}
\hypersetup{breaklinks=true,colorlinks=true,plainpages=true,linktocpage=true,linkcolor=MyDarkBlue,citecolor=MyDarkGreen,urlcolor=MyDarkRed,pdfborder={0 0 0},%
pdfauthor={Sang-Kil Son, Robert Thiele, Zoltan Jurek, Beata Ziaja, and Robin Santra},%
pdftitle={Quantum-mechanical calculation of ionization potential lowering in dense plasmas}%
}

\begin{document}
\begin{CJK*}{UTF8}{}


\title{Quantum-mechanical calculation of ionization potential lowering in dense plasmas}


\author{Sang-Kil Son \CJKfamily{mj}(손상길)}
\email{sangkil.son@cfel.de}
\affiliation{Center for Free-Electron Laser Science, DESY, Notkestrasse 85, 22607 Hamburg, Germany}
\affiliation{The Hamburg Centre for Ultrafast Imaging, Luruper Chaussee 149, 22761 Hamburg, Germany}

\author{Robert Thiele}
\email{robert.thiele@desy.de}
\affiliation{Center for Free-Electron Laser Science, DESY, Notkestrasse 85, 22607 Hamburg, Germany}
\affiliation{The Hamburg Centre for Ultrafast Imaging, Luruper Chaussee 149, 22761 Hamburg, Germany}

\author{Zoltan Jurek}
\affiliation{Center for Free-Electron Laser Science, DESY, Notkestrasse 85, 22607 Hamburg, Germany}
\affiliation{The Hamburg Centre for Ultrafast Imaging, Luruper Chaussee 149, 22761 Hamburg, Germany}

\author{Beata Ziaja}
\affiliation{Center for Free-Electron Laser Science, DESY, Notkestrasse 85, 22607 Hamburg, Germany}
\affiliation{The Hamburg Centre for Ultrafast Imaging, Luruper Chaussee 149, 22761 Hamburg, Germany}
\affiliation{Institute of Nuclear Physics, Polish Academy of Sciences, Radzikowskiego 152, 31-342 Krak{\'o}w, Poland}

\author{Robin Santra}
\affiliation{Center for Free-Electron Laser Science, DESY, Notkestrasse 85, 22607 Hamburg, Germany}
\affiliation{The Hamburg Centre for Ultrafast Imaging, Luruper Chaussee 149, 22761 Hamburg, Germany}
\affiliation{Department of Physics, University of Hamburg, Jungiusstrasse 9, 20355 Hamburg, Germany}

\date{\today}

\begin{abstract}
The charged environment within a dense plasma leads to the phenomenon of ionization potential depression (IPD) for ions embedded in the plasma. Accurate predictions of the IPD effect are of crucial importance for modeling atomic processes occurring within dense plasmas. Several theoretical models have been developed to describe the IPD effect, with frequently discrepant predictions. Only recently, first experiments on IPD in Al plasma have been performed with an x-ray free-electron laser (XFEL), where their results were found to be in disagreement with the widely-used IPD model by Stewart and Pyatt. Another experiment on Al, at the Orion laser, showed disagreement with the model by Ecker and Kr\"oll. This controversy shows a strong need for a rigorous and consistent theoretical approach to calculate the IPD effect. Here we propose such an approach: a two-step Hartree-Fock-Slater model. With this parameter-free model we can accurately and efficiently describe the experimental Al data and validate the accuracy of standard IPD models. Our model can be a useful tool for calculating atomic properties within dense plasmas with wide-ranging applications to studies on warm dense matter, shock experiments, planetary science, inertial confinement fusion and studies of non-equilibrium plasmas created with XFELs.
\end{abstract}


\maketitle
\end{CJK*}

\section{Introduction}\label{sec:intro}

The dense plasma state is a common phase of matter in the universe and can be found in all types of stars~\cite{Taylor94} and within giant planets~\cite{Chabrier09,Helled11}. 
Dense plasmas are commonly created during  experiments involving high-power light sources such as, e.g., the National Ignition Facility~\cite{Moses09}, and recently developed x-ray free-electron lasers (XFELs) LCLS~\cite{Emma10} and SACLA~\cite{Ishikawa12}.  
In dense plasmas, free electrons stay in the close vicinity of ions. 
The ions then cannot any longer be treated as isolated species, as the screening by the dense environment shifts their atomic energy levels, leading to a reduction of the ionization potentials. 
This effect is known as ionization potential depression (IPD).  
Quantitative predictions of this effect are of crucial importance for a correct understanding and accurate modeling of any atomic processes occurring within a dense plasma environment~\cite{WDM2011}, i.e., for studies on warm dense matter~\cite{Glenzer09,Drake09}, shock experiments~\cite{Lee09,Garcia-Saiz08}, planetary science~\cite{Knudson12,Nettelmann08}, inertial confinement fusion~\cite{Glenzer10,Lindl95} and studies of non-equilibrium plasmas created with XFELs~\cite{Corkum08,Faustlin10}.  
Several theoretical models have been developed to describe the IPD effect. 
An early development was the model proposed by Ecker and Kr\"oll (EK)~\cite{Ecker63} for strongly coupled plasma, later extended to the weakly coupled regime by Stewart and Pyatt (SP)~\cite{Stewart66} (for more examples, see Ref.~\cite{Murillo98}). 
However, until recently there were no experimental data available to verify the accuracy of these models whose predictions sometimes differed extensively.       

First experiments on the screening effect of plasma on atoms embedded in the plasma have been performed at LCLS~\cite{Vinko12,Ciricosta12,Cho12}. 
XFELs provide radiation of extremely high peak brightness and pulse duration shorter or comparable with the characteristic times of the electron and ion dynamics within irradiated systems. 
The dense electronic systems can quickly thermalize via electron--electron collisions and impact ionization processes~\cite{Faustlin10}. 
Because of the ultrashort pulse duration (typically tens of femtoseconds), only a thermalized electron plasma is probed while the ions still remain cold. 
This provides access to the properties of a solid-density material at temperature of $10^5$--$10^6$~K ($\approx 10$--100~eV).
Specifically, the experiments in Refs.~\cite{Vinko12,Ciricosta12,Cho12} measured  $K$-edge thresholds and $K\alpha$ emission from solid-density aluminum (Al) plasma. 
They have been followed by another experiment at the high-power Orion laser~\cite{Hoarty13,Hoarty13a}. 
This experiment investigated $K$-shell emissions from hot dense Al plasma. 
Both experimental teams tried to describe their findings with the EK and SP models. 
In the first experiment a disagreement of the measured $K$-edges with the extensively used SP model was claimed. 
A modified EK model was proposed to fit the experimental data~\cite{Ciricosta12,Preston13}. 
However, the data from the experiment on hot dense Al plasma~\cite{Hoarty13} could only be described with the SP model. 
The EK model was found to be in clear disagreement with these data.  

This controversy shows a strong need for a rigorous and consistent theoretical approach able to calculate the IPD effect for plasmas in different coupling regimes. 
Here we propose such an approach: a two-step Hartree-Fock-Slater (HFS) method. 
This model derives the electronic structure of an ion embedded in the electron plasma from the finite-temperature approach~\cite{Mermin63}, assuming thermalization of bound electrons within the free-electron plasma.  
It can also treat individual electronic configurations of plasma ions, which enables a description of discrete transitions.  
In this paper, we demonstrate that this model successfully describes laser-irradiated Al solids under the conditions of the LCLS~\cite{Vinko12,Ciricosta12,Cho12} and Orion laser~\cite{Hoarty13,Hoarty13a} experiments.
Furthermore, we gain an improved understanding of the validity of the widely-used EK and SP models. 

\section{Two-step Hartree-Fock-Slater model}\label{sec:theory}

In the first step, we apply the thermal HFS approach for a given finite temperature.
Here we assume that the electrons are fully thermalized.
In general, XFEL radiation induces non-equilibrium dynamics of electrons, for instance, in carbon-based materials exposed to hard X-rays~\cite{Hau-Riege13}.
For the solid-density Al plasma studied in the recent experiment~\cite{Vinko12}, the incident photon energy is near the ionization threshold, ejecting electrons with low kinetic energy.
Due to the high density and low kinetic energy, electron cross sections are large, so that electrons equilibrate rapidly within the ultrashort pulse duration.
For example, in the Al plasma considered here, the energy of a photoelectron is less than 270~eV and the highest energy of an Auger electron is about 1.4~keV.
With a kinetic energy in this regime, the estimated time scale of electron thermalization is a few femtoseconds for diamond~\cite{Ziaja05} and is expected to be shorter for solid Al due to higher impact ionization cross sections.
Therefore, we assume that the electrons are thermalized within the pulse duration of tens of femtoseconds.

The standard Hartree-Fock and Hartree-Fock-Slater (HFS) approaches~\cite{Slater51,Herman63} treat electronic structure at zero temperature ($T=$0~eV). 
They use the Ritz variational principle for the ground-state energy. 
Recently, Thiele~\textit{et al.}~\cite{Thiele12} proposed an extension of the standard HFS model, including plasma environment effects through the Debye screening (see also Refs.~\cite{Saha02,Mukherjee02,Das11}). 
However, this model is not applicable to plasmas with low temperature, where the Debye screening approximation breaks down~\cite{Murillo98}. 
Also, it is intrinsically inconsistent to combine the $T=0$~eV approach with Debye screening for a plasma with a non-zero electron temperature. 
To overcome this inconsistency, the electronic structure has to be derived from a finite-temperature approach. 
Such a finite-temperature Hartree-Fock approach was proposed by \citet{Mermin63}. 

Here we use the average-atom model~\cite{Rozsnyai72}, which is a variant of the finite-temperature approach. 
Basically, it predicts average orbital properties and occupation numbers at a given temperature.
There have been various implementations of the average-atom approach.
Depending on their treatment of the electronic structure of atoms, they can be categorized as quantum mechanical approaches, such as the Hartree-Fock-Slater (HFS) method or local density approximation (LDA)~\cite{Rozsnyai72,Liberman79,Blenski95,Johnson06,Sahoo08,Johnson12,Cauble84,Davis90,Piron11,Starrett13,Pain06,Pain06a}, or semi-classical approaches, such as the Thomas-Fermi (TF) method~\cite{Feynman49,Rozsnyai72,Crowley90,Yuan02}.
There are also implementations with simplified super-configurations~\cite{Blenski97,Blenski97a,Faussurier99,Pain06,Pain06a} and with the screened hydrogenic model~\cite{More82,Faussurier97,Faussurier97a,Faussurier99}.
The atomic potential within a plasma is usually based on the muffin-tin approximation~\cite{Liberman79,Blenski95,Johnson06,Sahoo08,Johnson12,Rozsnyai72,Yuan02} or an extended model including ion--ion correlation~\cite{Cauble84,Davis90,Crowley90,Starrett13}.
These average-atom models have been applied to calculate physical quantities within plasmas, such as lowering of the ionization energy~\cite{Davis90}, photoabsorption processes~\cite{Blenski97,Johnson06}, and scattering processes~\cite{Johnson12,Johnson13}.
Our average-atom implementation presented here is based on the quantum mechanical approach with the muffin-tin approximation~\cite{Blenski95,Johnson06,Sahoo08,Johnson12}, but it benefits from a numerical grid technique, which will be discussed in detail later.

The major distinction of our proposed method from previous average-atom models is not only the different numerical method.
In this paper, we will develop a simple model to retrieve more complete information from the average-atom approach.
We propose a two-step model: an average-atom calculation as the first step and a fixed-configuration calculation as the second step.
From the average-atom approach, we obtain a grand-canonical ensemble at a given temperature with a simple electronic mean field associated with all possible configurations.
Using this information, we then calculate improved mean fields for selected configurations of interest.
Note that our two-step model is relatively inexpensive, in comparison with polyatomic density-functional calculations, and is easily applicable to any atomic species.
This two-step model is described in detail in the three following subsections.

\subsection{Hartree-Fock-Slater calculation with a muffin-tin-type potential}\label{sec:HFS}

\begin{figure}
\centering
\includegraphics[width=\figurewidth]{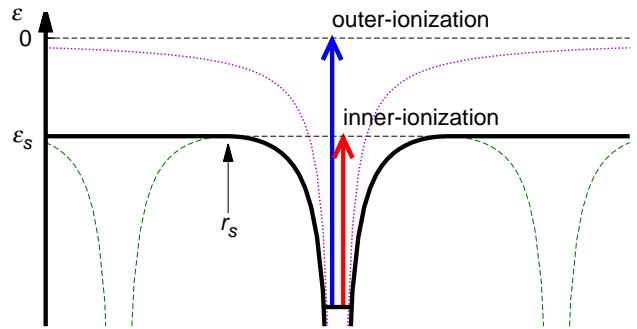}
\caption{\label{fig:potential_solid}%
Schematic diagram of an atomic model in a solid or a plasma.
The purple dotted curve is the isolated atomic potential, the green dashed curve is the crystalline potential, and the black thick curve is the muffin-tin-type potential.
The muffin-tin flat potential is denoted as $\varepsilon_s$ and the Wigner-Seitz radius is $r_s$.
}
\end{figure}

To describe the electronic structure in a solid or a plasma, we employ a muffin-tin-type potential~\cite{Slater37} as depicted in Fig.~\ref{fig:potential_solid}.
Influenced by the free electrons and neighboring ions, the atomic potential is lowered in comparison to that in an isolated atom.
%
The sphere surrounding an atom is defined by the Wigner-Seitz radius, $r_s$.
If the solid consists of a single atomic species, then $r_s = ( 3 / 4 \pi n_i )^{1/3}$, where $n_i$ is the number density of ions in the solid.
Here we assume that the positions of the ions are fixed.
Therefore the Wigner-Seitz radius does not change during the calculation.
We assume charge neutrality such that the ionic charge density outside the Wigner-Seitz sphere is compensated by the electron density.
The net charge inside the Wigner-Seitz sphere is also zero on average, so that the potential outside is constant.
We use this muffin-tin-type model for all our calculations.

Our implementation of the muffin-tin potential differs from the original model suggested by Slater~\cite{Slater37} and its quantum-mechanical implementations~\cite{Blenski95,Johnson06,Sahoo08,Johnson12}.
First, the constant potential outside the atomic sphere is self-consistently calculated in our model, whereas it is set to zero in other previous implementations~\cite{Slater37,Blenski95,Johnson06,Sahoo08,Johnson12}.
We refer to this constant potential as the muffin-tin flat potential, $\varepsilon_s$.
Second, we calculate both bound- and continuum-state wave functions with the same atomic potential, using numerical grids with a sufficiently large radius far from $r_s$.
This makes our method distinct from other implementations where a continuum state outside $r_s$ is usually given as a plane wave and special boundary conditions for both bound and continuum states are required.

With the muffin-tin flat potential, we may consider different ionization processes in a solid or a cluster~\cite{Last99,Krainov02}.
The ionization energy is defined as the energy needed to transfer an electron to the continuum level located at $\varepsilon=0$, which corresponds to the binding energy measured with photoelectron spectroscopy. 
In a solid or a cluster, this process would be called \emph{outer-ionization}. 
On the other hand, there is already a continuum of states starting at $\varepsilon_s$, when the muffin-tin-type potential is imposed. 
It defines excitation into the continuum for $\varepsilon_s \leq \varepsilon \leq 0$, which would be called \emph{inner-ionization}. 
In metals like aluminum, the conduction band can be described by this continuum above $\varepsilon_s$ and inner-ionization is a transfer of an electron bound to an atom (narrow band) into the conduction band. 
Figure~\ref{fig:potential_solid} depicts schematically outer-ionization and inner-ionization processes for the muffin-tin-type potential.

We solve the effective single-electron Schr{\"o}dinger equation with the muffin-tin-type potential (atomic units are used unless specified otherwise),
\begin{equation}\label{eq:SE}
\left[ -\frac{1}{2} \nabla^2 + V(\mathbf{r}) \right] \psi(\mathbf{r}) = \varepsilon \psi(\mathbf{r}),
\end{equation}
where the potential is the Hartree-Fock-Slater (HFS) potential inside $r_s$ and is constant outside $r_s$,
%
\begin{equation}\label{eq:potential}
V(\mathbf{r}) = 
\begin{cases}
{\displaystyle - \frac{Z}{r} + \int_{r' \leq r_s} d^3 r' \; \frac{ \rho(\mathbf{r}') }{ | \mathbf{r} - \mathbf{r}' | } + V_\text{x}[ \rho(\mathbf{r}) ] } &  \text{for }r \leq r_s, 
\\
{\displaystyle V(r_s)} &  \text{for }r > r_s,
\end{cases}
\end{equation}
%
where $Z$ is the nuclear charge, $\rho(\mathbf{r})$ is the electronic density, and $V_\text{x}$ is the Slater exchange potential, 
\begin{equation}\label{eq:Slater}
V_\text{x}[ \rho(\mathbf{r}) ] = - \frac{3}{2} \left[ \frac{3}{\pi} \rho(\mathbf{r}) \right]^{1/3}.
\end{equation}
We use a spherically symmetric electronic density, $\rho(\mathbf{r}) \rightarrow \rho(r)$, so $V(\mathbf{r})$ is also spherically symmetric.
The potential for $r > r_s$ is given by the constant value of $V(r_s)$, fulfilling the continuity condition at the boundary $r = r_s$.
This constant potential defines the muffin-tin flat potential, $\varepsilon_s = V(r_s)$.

For an isolated atom without the plasma, we use the original HFS potential, which is Eq.~\eqref{eq:potential} without applying the Wigner-Seitz radius and the muffin-tin flat potential, 
\begin{equation}\label{eq:unscreened}
V_\text{atom}(\mathbf{r}) = 
{\displaystyle - \frac{Z}{r} + \int d^3 r' \; \frac{ \rho(\mathbf{r}') }{ | \mathbf{r} - \mathbf{r}' | } + V_\text{x}[ \rho(\mathbf{r}) ] }.
\end{equation}
In contrast to plasmas and solids, the long-range potential in the isolated atom is governed by the Coulomb potential [$ = - (Q + 1) / r$] where $Q$ is the charge of the atomic system.
However, it is well known that the original HFS potential does not have this proper asymptotic behavior due to the self-interaction term~\cite{Parr89}.
To obtain the proper long-range potential for both occupied and unoccupied orbitals, we apply the Latter tail correction~\cite{Latter55}.
Thus, the unscreened HFS potential for an isolated atom is given by Eq.~\eqref{eq:unscreened} and replaced by $-(Q+1)/r$ only when $V_\text{atom}(\mathbf{r}) > -(Q+1)/r$.

The orbital wave function $\psi(\mathbf{r})$ is expressed in terms of the product of a radial wave function and a spherical harmonic,
\begin{equation}\label{eq:wf}
\psi_{nlm}(\mathbf{r}) = \frac{u_{nl}(r)}{r} Y^m_l(\theta,\phi),
\end{equation}
where $n$, $l$, and $m$ are the principal quantum number, the orbital angular momentum quantum number, and the associated projection quantum number, respectively.
The radial wave function $u_{nl}(r)$ is solved by the generalized pseudospectral method~\cite{Yao93a,Tong97a}.
Plugging Eq.~\eqref{eq:wf} into Eq.~\eqref{eq:SE}, we obtain the radial Schr\"odingier equation for a given $l$,
\begin{equation}\label{eq:radial_SE}
\left[ -\frac{1}{2} \frac{d^2}{dr^2} + \frac{l(l+1)}{2 r^2} + V(r) \right] u_{nl}(r) = \varepsilon_{nl} u_{nl}(r).
\end{equation}
The Hamiltonian and radial wave function in Eq.~\eqref{eq:radial_SE} are discretized on a nonuniform grid for $0 \leq r \leq r_\text{max}$.
Note that the boundary conditions are $u_{nl}(0) = u_{nl}(r_\text{max}) = 0$, and no additional boundary condition at $r_s$ is imposed.
After diagonalizing the discretized Hamiltonian matrix, one obtains not only bound states ($\varepsilon < \varepsilon_s$) but also a discretized \textit{pseudocontinuum} ($\varepsilon \geq \varepsilon_s$).
With a sufficiently large radius $r_\text{max}$, the distribution of these pseudocontinuum states becomes dense enough to imitate continuum states~\cite{Chu77,Santra02,Chu04}.

\begin{figure}
\centering
\includegraphics[width=\figurewidth]{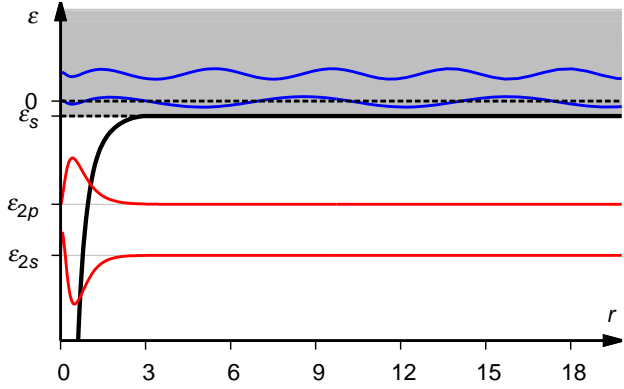}
\caption{\label{fig:Al_pot_wf}%
Plots of bound- and continuum-state energies of Al at zero temperature, obtained by diagonalization of the discretized Hamiltonian matrix.
Some of bound- and continuum-state wave functions are also plotted.
The maximum radius is $r_\text{max} = 100$~a.u.\ used in the calculation.
}
\end{figure}

Figure~\ref{fig:Al_pot_wf} shows some pseudocontinuum states as well as bound states ($2s$ and $2p$) of Al at zero temperature.
The dense horizontal lines above $\varepsilon_s$ are all pseudocontinuum states.
Here we use $r_\text{max}=100$~a.u., which is much larger than $r_s=2.99$~a.u.\ from the Al solid density (2.7~g/cm$^3$).
The number of grid points for $r$ is 200 and the number of partial waves is 31 ($0 \leq l \leq 30)$.
With these computational parameters, we obtain 6200 radial eigenstates.
In the metallic Al case, only 3 states ($1s$, $2s$, and $2p$) are bound states.
All other eigenstates constitute a pseudocontinuum.
We keep these computational parameters for all calculations throughout the paper.

The numerical grid technique we use here attains advantageous simplicity in continuum-state calculations.
One can transform the integration over positive energy into a summation over discrete states.
There is no separation of the inside and outside regions, and therefore, no boundary condition at the Wigner-Seitz radius is needed.
In contrast, other implementations of the muffin-tin model involve special boundary conditions at $r_s$.
For example, \citet{Johnson06,Johnson12} used the condition that the wave functions in the inner sphere are continuously connected to those in the outside region, and \citet{Sahoo08} enforced the derivative of the wave function to vanish at the Wigner-Seitz radius.
Note that different boundary conditions at $r_s$ lead to different electronic structures as pointed out in Ref.~\cite{Johnson12}.

\subsection{The first step: average-atom calculation}\label{sec:first-step}
The first step of our two-step HFS approach is an average-atom model calculation with the muffin-tin-type potential in Eq.~\eqref{eq:potential}.
%
We treat the electronic system using a grand-canonical ensemble at a finite temperature $T$ (in units of energy). 
The electronic density $\rho(\mathbf{r}, T)$ is then constructed by
\begin{align}\label{eq:total_density}
\rho(\mathbf{r}, T) 
&= \sum_p \left| \psi_p(\mathbf{r}) \right|^2 \tilde{n}_p(\mu, T),
\end{align}
where $p$ indicates the one-particle state index, i.e., $p = (n, l, m, m_s)$ where $m_s$ is the spin quantum number, and $p$ runs over all bound and continuum states.
Here $\{ \tilde{n}_p(\mu, T) \}$ are fractional occupation numbers according to the Fermi-Dirac distribution with a chemical potential $\mu$,
\begin{equation}
\tilde{n}_p(\mu, T) = \frac{ 1 }{ e^{( \varepsilon_p - \mu ) / T} + 1 },
\end{equation}
where $\varepsilon_p$ is the orbital energy for a given spin-orbital $p$.
The average number of electrons, $N_\text{elec}$, within the Wigner-Seitz sphere,
\begin{equation}\label{eq:N_elec}
N_\text{elec} = \int_{r \leq r_s} d^3 r \; \rho(\mathbf{r}, T),
\end{equation}
is fixed to $N_\text{elec} = Z$ to ensure charge neutrality. 
This condition serves as a constraint to determine the chemical potential at the given temperature~\cite{Johnson06,Sahoo08,Johnson12}. 
In order to determine $\mu$, one must find the root of the following equation,
\begin{equation}\label{eq:f(mu)}
N_\text{elec} - \sum_p \left( \int_{r \leq r_s} d^3 r \; \left| \psi_p(\mathbf{r}) \right|^2 \right) \tilde{n}_p(\mu, T) = 0.
\end{equation}
With $\mu$ obtained from Eq.~\eqref{eq:f(mu)}, $\rho(\mathbf{r}, T)$ is constructed from Eq.~\eqref{eq:total_density}.
With $\rho(\mathbf{r}, T)$, the updated atomic potential, as well as $\varepsilon_s$, is obtained from Eq.~\eqref{eq:potential}.
Then, orbitals $\{ \psi_p(\mathbf{r}) \}$ and orbital energies $\{ \varepsilon_p \}$ are calculated, using the new potential.
Again, a new $\mu$ is obtained from Eq.~\eqref{eq:f(mu)}.
This self-consistent field (SCF) procedure is performed until the results converge.
Note that there are only three input parameters in the calculation: element species ($Z$), temperature ($T$), and solid density via the Wigner-Seitz radius ($r_s$).
All other quantities such as orbitals, orbital energies, $\rho$, $\mu$, and $\varepsilon_s$ are determined self-consistently.

Regarding the exchange potential at a finite temperature, various implementations have been proposed~\cite{Rozsnyai72,Shalitin73,Perrot84,Pittalis11,Karasiev12,Pribram-Jones14}, but no unanimous expression has been identified.
\citet{Perrot84} proposed a parameterization of the thermal exchange potential based on the local density approximation (LDA).
\citet{Rozsnyai72} proposed an interpolation between the zero-temperature Slater potential and the high temperature limit.
In the present calculations, we use the same potential as used in the zero-temperature calculation given by Eq.~\eqref{eq:Slater}.
Note that our approach can be easily combined with any type of exchange potential.
We will discuss the dependence on different thermal exchange potentials in Sec.~\ref{sec:results}.

%

\subsection{The second step: fixed-configuration calculation}\label{sec:second-step}
The second step in our two-step HFS approach is a fixed configuration calculation for bound electrons in the presence of the free-electron density. 
Within the average-atom model, one cannot obtain orbital energies of individual electronic configurations associated with different charge states.
Instead, orbital energies in the average-atom model represent averaged quantities for an averaged configuration with fractional occupational numbers.
However, in a fluorescence experiment, for instance, one can see discrete transition lines corresponding to individual charge states~\cite{Vinko12}, which are not accessible within the average-atom model.
In order to describe individual electronic configurations within a plasma environment, we propose a fixed-configuration scheme. 

With the grand-canonical ensemble, one can calculate the probability distribution~\cite{Pei00} of all possible bound-state configurations (see Appendix~\ref{appendix1} for details),
\begin{equation}\label{eq:prob_bound_config}
P_{[n_b]} = \prod_{b}^\text{bound}
\frac{ e^{ -( \varepsilon_{b} - \mu ) n_b / T } }{ 1 + e^{ -( \varepsilon_{b} - \mu ) / T } },
\end{equation}
where $[n_b] = ( n_1, \cdots, n_B )$ indicates the fixed bound-state configuration, and $B$ is the number of bound one-electron states.
Here $b$ runs over all bound states $(1 \leq b \leq B)$ and $n_b$ is an integer occupation number (0 or 1) in the bound-electron configuration.
The probability of finding the charge state $Q$ is given by the sum of all associated bound-state configurations,
\begin{equation}\label{eq:prob_charge}
P_Q = \sum_{[n_b]}^{Q} P_{[n_b]}.
\end{equation}
Here $[n_b]$ runs over all possible bound-state configurations satisfying $\sum_b^\text{bound} n_b = Z - Q$.

From the probability distributions in Eqs.~\eqref{eq:prob_bound_config} and \eqref{eq:prob_charge}, one can choose one bound-electron configuration from the grand-canonical ensemble and perform a single SCF calculation with this fixed configuration.
For example, it is possible to choose the most probable configuration associated with the most probable charge state.
The $K\alpha$ transition energy calculated from this configuration gives one discrete line in the x-ray emission spectrum.
Different configurations contribute to different transition lines, so measurement of these lines maps out the distribution of all possible configurations and charge states.

Here we focus on individual bound-electron configurations.
Even though the bound-state electronic structure is influenced by the presence of the plasma electrons, we assume that it is not sensitive to detailed free-electron configurations in the plasma.
Therefore, once we choose one bound-electron configuration, we calculate the free-electron density as an average of all possible free-electron configurations for the given bound-electron configuration (see Appendix~\ref{appendix2}),
\begin{equation}\label{eq:rho_f}
\rho_f(\mathbf{r},T) = \sum_{p}^\text{continuum} \left| \psi_{p}(\mathbf{r}) \right|^2 \tilde{n}_p(\mu,T),
\end{equation}
which turns out to be independent of the bound-electron configuration selected.
This free-electron density is self-consistently obtained in the first step and is kept fixed in the second step.
The bound-electron density is constructed with a fixed electron configuration,
\begin{equation}
\rho_b(\mathbf{r}) = \sum_b^\text{bound} \left| \psi_b(\mathbf{r}) \right|^2 n_b.
\end{equation}
Then, the total electron density is constructed as the sum of the bound and free-electron densities,
\begin{equation}\label{eq:rho}
\rho(\mathbf{r}) = \rho_b(\mathbf{r}) + \rho_f(\mathbf{r},T).
\end{equation}
%
%
With this total electron density, we perform a HFS calculation using a microcanonical ensemble.
In this case, $\rho_b$ is self-consistently updated, whereas $\rho_f$ remains fixed during the SCF procedure. 
This approach allows the bound electrons in a given configuration to adjust to the presence of the plasma electrons.

\section{Applications of two-step HFS model to Al plasmas}\label{sec:results}
In this section, we apply the two-step HFS model to laser-irradiated Al solid~\cite{Vinko12,Ciricosta12,Hoarty13}. 
%
The two-step procedure is carried out as follows.
First, we perform a finite-temperature HFS calculation to determine the temperature needed to achieve a certain average charge state within the plasma. 
We then use the free-electron density $\rho_f(\mathbf{r},T)$ obtained from the finite-temperature calculation to perform a fixed-configuration HFS calculation for the given charge state to determine the energy of the $1s$ orbital and the energy of the energetically lowest $p$ orbital above $2p$ (always referred to as $3p$, whether it is bound or not). 
%
Both the first and second steps of our two-step HFS model are implemented as an extension of the \textsc{xatom} toolkit~\cite{xatom,Son11a}, which can calculate any atomic element and any electronic configuration within a non-relativistic framework.
All calculations are performed with the computational parameters stated in Sec.~\ref{sec:first-step} and are fully converged.

\subsection{Al plasmas at low temperature calculated in the first step}

\begin{table}
\caption{\label{table:Al}%
Electronic structure of Al metal and Al plasma.
All energies in our present calculations are subtracted by $\varepsilon_s$, in order to compare with other theoretical data.
Al solid density is $n_i^\circ=2.7$~g/cm$^3$ and $\bar{Q}$ is the average charge state.
Energies are in eV.
}
\begin{ruledtabular}
\begin{tabular}{crrrr}
& \multicolumn{2}{c}{$T=10$~eV, $n_i=n_i^\circ$} & \multicolumn{2}{c}{$T=5$~eV, $n_i=0.1n_i^\circ$}
\\
Level		& Present 	& Ref.~\cite{Sahoo08}	& Present 	& Ref.~\cite{Johnson06} \\
\hline
$1s$		& $-1530.1$	& $-1495.0$				& $-1547.3$	& $-1501.8$	\\
$2s$		& $-102.5$ 	& $-101.2$				& $-119.6$	& $-108.3$	\\
$2p$		& $-64.9$ 	& $-63.9$				& $-82.0$	& $-71.0$	\\
$3s$		& --		& $-6.3$				& $-8.6$	& $-7.0$	\\
$3p$		& --		& --					& $-2.5$	& $-1.5$	\\
$\mu$		& $-1.5$	& --					& $-11.1$	& $-10.4$	\\
\hline
$\bar{Q}$	& $+3.01$	& $+2.38$				& $+1.34$	& $+1.49$
\end{tabular}
\end{ruledtabular}
\end{table}


%
In order to benchmark our calculations, we apply our average-atom model (the first step of our two-step approach) to Al solid ($T=0$~eV) and low-temperature Al plasma ($T \leq 10$~eV).
In our average-atom model, we self-consistently determine all orbital energies, the muffin-tin flat potential, and the chemical potential.
The Fermi level is the position of the chemical potential $\mu$ at $T=0$~eV.
The muffin-tin flat potential $\varepsilon_s$ represents the lower limit of the delocalized states, corresponding to the lowest occupied state in the conduction band.
Therefore, if one defines the Fermi energy relative to the beginning of the conduction band, it is given by $\varepsilon_F = \mu - \varepsilon_s$ in our calculation.
For $T>0$~eV, the $K$-shell inner-ionization energy is defined by the difference between the muffin-tin flat potential and the $1s$ orbital energy, $E_K = \varepsilon_s - \varepsilon_{1s}$, employing the zeroth-order approximation for the HFS energy, which is similar to Koopmans' theorem for the Hartree-Fock method. 
For the Al solid at $T=0$~eV, the inner-ionization energy is given by $E_K = \mu - \varepsilon_{1s}$, because orbitals below the Fermi level are fully occupied and there are no transitions into those orbitals.
As the temperature increases, the chemical potential becomes lower than the muffin-tin flat potential and the occupation numbers in the continuum states follow the thermal Boltzmann distribution for the given plasma temperature. 
For instance, the occupation number in $3s$ and $3p$ is $\sim$0.46 at $T=10$~eV and $\sim$0.18 at $T=30$~eV. 
In this way, for $T>0$~eV the continuum states above the muffin-tin flat potential become available for electronic transitions.  

Taking into account the HFS approximation and the muffin-tin approximation, our calculations provide reasonable electronic structures for metallic states.
For Al at $T=0$~eV and solid density (2.7~g/cm$^3$), the average charge $\bar{Q}$ is $+3$, indicating that only 10 electrons are bound as $1s^2$, $2s^2$, and $2p^6$. 
The other 3 electrons, which would be $3s^2$$3p^1$ in an isolated atom, are then already in the continuum, i.e., within the conduction band.
Therefore, our model can mimic the electronic structure of the metal.
With our method, we found $\varepsilon_F=8.0$~eV. 
The experimental Fermi energy is 11.7~eV~\cite{Ashcroft76}.
Our method calculates $E_K = 1538.1$~eV at $T=0$~eV, while the experimental binding energy of the $K$-shell relative to the Fermi level is 1559.6~eV~\cite{Thompson01}.

For $T > 0$~eV, Table~\ref{table:Al} compares our results with available theoretical data.
All energies in our calculations are subtracted by $\varepsilon_s$ in order to compare with previous calculations~\cite{Johnson06,Sahoo08} where $\varepsilon_s$ is set to zero.
We found that for $T \leq 10$~eV at solid density the average charge of Al is $+3$, and $M$-shells ($3s$ and $3p$) are not bound.
Our finding agrees with the comment in Ref.~\cite{Johnson12}, but disagrees with the results in Ref.~\cite{Sahoo08}, where $3s$ is bound in this low temperature regime.
In Table~\ref{table:Al}, we list energy levels of the Al plasma at $T=10$~eV in comparison with Ref.~\cite{Sahoo08}.
The different prediction for $M$-shell binding is ascribed to the different boundary condition as discussed in Ref.~\cite{Johnson12}.
At $T=5$~eV and density of 0.27~g/cm$^3$, we compare our results with Ref.~\cite{Johnson06}.
The discrepancy in this case is entirely due to the different exchange potential.
If we use the same LDA potential as used in Ref.~\cite{Johnson06} ($V_\text{x}^\text{LDA} = \frac{2}{3} V_\text{x}$), then we obtain the same results for all energy levels and the averaged charge state.

\subsection{Connection between the first and second steps: bound-electron configuration and free-electron density}

\begin{figure}
\centering
\includegraphics[width=\figurewidth]{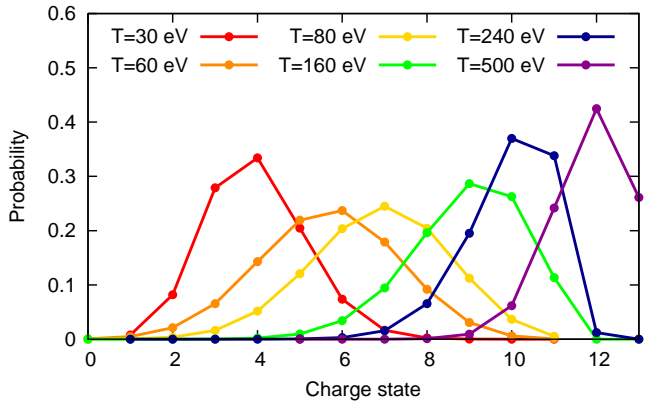}
\caption{Probability distribution of charge states for different temperatures calculated in the first step of our two-step HFS model.}
\label{fig:Al_population}
\end{figure}

\begin{table}
\caption{\label{table:config_probability}
Probability distribution of bound-electron configurations at $T=80$~eV.
Configurations are listed when their probability is greater than 0.01, and the probability is calculated from the first step. 
$K$-shell ionization energy ($E_K$) and $K\alpha$ transition energy ($E_{K\alpha}$) are calculated from the second step of our two-step HFS model.
$E_K$ and $E_{K\alpha}$ are in eV.
}
\begin{ruledtabular}
\begin{tabular}{ccccc}
$Q$	& Configuration					& Probability	& $E_K$		& $E_{K\alpha}$ \\
\hline
+5	& $1s^2 2s^1 2p^4 3s^0 3p^1$	& 0.0193 		& 1618.3	& 1497.7 \\
	& $1s^2 2s^2 2p^3 3s^0 3p^1$	& 0.0187 		& 1623.1	& 1500.3 \\
	& $1s^2 2s^2 2p^4 3s^0 3p^0$	& 0.0174		& 1578.7	& 1486.7 \\
\hline
+6	& $1s^2 2s^1 2p^3 3s^0 3p^1$	& 0.0376		& 1658.1	& 1511.6 \\
	& $1s^2 2s^1 2p^4 3s^0 3p^0$	& 0.0349		& 1618.3	& 1497.7 \\
	& $1s^2 2s^2 2p^3 3s^0 3p^0$	& 0.0339		& 1623.1	& 1500.3 \\
	& $1s^2 2s^2 2p^2 3s^0 3p^1$ 	& 0.0205		& 1663.5	& 1514.5 \\
	& $1s^2 2s^1 2p^3 3s^1 3p^0$ 	& 0.0139		& 1656.0	& 1511.3 \\
\hline
+7	& $1s^2 2s^1 2p^3 3s^0 3p^0$	& 0.0681		& 1666.3	& 1512.8 \\
	& $1s^2 2s^1 2p^2 3s^0 3p^1$	& 0.0413		& 1705.4	& 1527.8 \\
	& $1s^2 2s^2 2p^2 3s^0 3p^0$	& 0.0371		& 1671.9	& 1515.8 \\
	& $1s^2 2s^0 2p^3 3s^0 3p^1$	& 0.0189		& 1699.3	& 1524.5 \\
	& $1s^2 2s^0 2p^4 3s^0 3p^0$	& 0.0175		& 1660.9	& 1509.9 \\
	& $1s^2 2s^1 2p^2 3s^1 3p^0$	& 0.0153		& 1705.4	& 1527.9 \\
	& $1s^2 2s^2 2p^1 3s^0 3p^1$	& 0.0120		& 1711.7	& 1531.2 \\
\hline
+8	& $1s^2 2s^1 2p^2 3s^0 3p^0$	& 0.0747		& 1718.7	& 1530.0 \\
	& $1s^2 2s^0 2p^3 3s^0 3p^0$	& 0.0342		& 1712.3	& 1526.7 \\
	& $1s^2 2s^1 2p^1 3s^0 3p^1$	& 0.0241		& 1758.5	& 1546.5 \\
	& $1s^2 2s^2 2p^1 3s^0 3p^0$	& 0.0217		& 1725.1	& 1533.4 \\
	& $1s^2 2s^0 2p^2 3s^0 3p^1$	& 0.0207		& 1751.6	& 1542.9 \\
\hline
+9	& $1s^2 2s^1 2p^1 3s^0 3p^0$	& 0.0437		& 1775.1	& 1549.6 \\
	& $1s^2 2s^0 2p^2 3s^0 3p^0$	& 0.0375		& 1768.0	& 1545.9 \\
	& $1s^2 2s^0 2p^1 3s^0 3p^1$	& 0.0121		& 1808.2	& 1564.1 \\
\hline
+10	& $1s^2 2s^0 2p^1 3s^0 3p^0$	& 0.0219		& 1827.4	& 1568.1 \\
	& $1s^2 2s^1 2p^0 3s^0 3p^0$	& 0.0106		& 1835.2	& 1572.1 \\
\end{tabular}
\end{ruledtabular}
\end{table}

As discussed in Sec.~\ref{sec:second-step}, we choose certain fixed configurations based on the probability distribution of charge states and bound-electron configurations.
Figure~\ref{fig:Al_population} shows the charge state distribution for $T=30$--500~eV, calculated using Eq.~\eqref{eq:prob_charge}.
As $T$ increases, the charge state distribution moves toward higher charge states, resulting in higher $\bar{Q}$.
%
From the first step, it is also possible to calculate probabilities for all possible bound-electron configurations associated with individual charge states.
For example, Table~\ref{table:config_probability} shows bound-electron configurations at $T=80$~eV, whose probability is greater than 0.01, calculated using Eq.~\eqref{eq:prob_bound_config}.
These probability distributions of charge states and bound-electron configurations provide detailed information about the ensemble and enable us to perform the second step of our two-step approach.
In Table~\ref{table:config_probability}, we also list $K$-shell ionization energies ($=\varepsilon_s - \varepsilon_{1s}$) and $K\alpha$ transition energies ($=\varepsilon_{2p} - \varepsilon_{1s}$), calculated from the second step.
Individual configurations provide different ionization energies and transition energies, which cannot be captured by averaged orbital energies from the average-atom approach only.
Note that the ground-state configuration is usually not the most probable configuration for given charge states, illustrating the importance of detailed electronic structures of individual configurations.
In our calculation, $3s$ and $3p$ are bound at $T=80$~eV, and they are included in the bound-electron configuration.
However, those $M$-shell electrons do not considerably alter the $1s$--$2p$ transition lines.
For example, $E_{K\alpha}=1512.8$~eV for Al$^{7+}$ $1s^2 2s^1 2p^3$ is similar to 1511.6~eV for Al$^{6+}$ $1s^2 2s^1 2p^3 3p^1$ and 1511.3~eV for Al$^{6+}$ $1s^2 2s^1 2p^3 3s^1$ (see Table~\ref{table:config_probability}).
To compare calculated $K\alpha$ lines with experimental results, it is plausible to assign them according to the super-configuration of $K$ and $L$-shells only, as suggested in Refs.~\cite{Vinko12,Cho12}.

\begin{figure}
\centering
\includegraphics[width=\figurewidth]{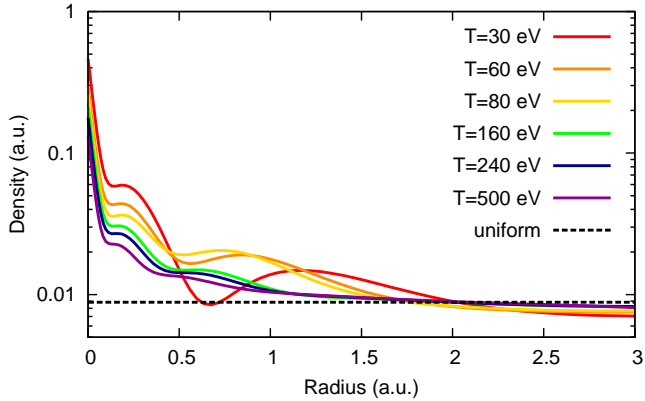}
\caption{\label{fig:Al_density}%
Free-electron density of aluminum plasma for different temperatures, obtained from the first step of our two-step model.
}
\end{figure}

The free-electron density is obtained from the first step of the two-step model.
Figure~\ref{fig:Al_density} shows the free-electron density $\rho_f(\mathbf{r},T)$ for different electronic temperatures ($T=30$--500~eV), calculated using Eq.~\eqref{eq:rho_f}.
The density plot is normalized such that the integration of the density within $r_s$ yields one.
This free-electron density is self-consistently optimized in the presence of the central nucleus and bound electrons, thus its distribution is highly non-uniform. 
As expected, the free-electron density tends to be more uniformly distributed within the Wigner-Seitz sphere at higher temperatures. 
For comparison, a constant and normalized density is also plotted with a dashed line in Fig.~\ref{fig:Al_density}.
The shape of the free-electron density at $T=30$~eV is attributed to the nodal structure of the $3p$ orbital in the continuum.

\subsection{Ionization potential depression in Al plasmas: LCLS experiment}

\begin{table}
\caption{\label{table:config}%
Average charge state $\bar{Q}$, the most probable charge state $Q_\text{mp}$ and the most probable bound-electron configuration $C_\text{mp}$ for a given temperature $T$ from the first step of two-step HFS calculation.
Note that $C_\text{mp}$ is the ground configuration for a given charge state, except Al$^{7+}$ at $T=80$~eV.
}
\begin{ruledtabular}
\begin{tabular}{cccc}
$T$ & $\bar{Q}$ & $Q_\text{mp}$ & $C_\text{mp}$
\\
\hline
10 & +3.01 & +3 & $1s^2 2s^2 2p^6$
\\
30 & +3.95 & +4 & $1s^2 2s^2 2p^5$
\\
40 & +4.83 & +5 & $1s^2 2s^2 2p^4$
\\
60 & +5.67 & +6 & $1s^2 2s^2 2p^3$
\\
80 & +6.87 & +7 & $1s^2 2s^1 2p^3$
\end{tabular}
\end{ruledtabular}
\end{table}

As shown in the previous subsection, we determine from the first step for a given temperature: a) probabilities of all individual electronic configurations associated with different charge states, and b) the free-electron density.  
For the LCLS conditions ($T=10$--80~eV) corresponding to the strongly and moderately coupled plasma regimes, the average charge state, the most probable charge state, and the most probable configuration of this charge state are listed in Table~\ref{table:config}.

\begin{figure}
\centering
\includegraphics[width=\figurewidth]{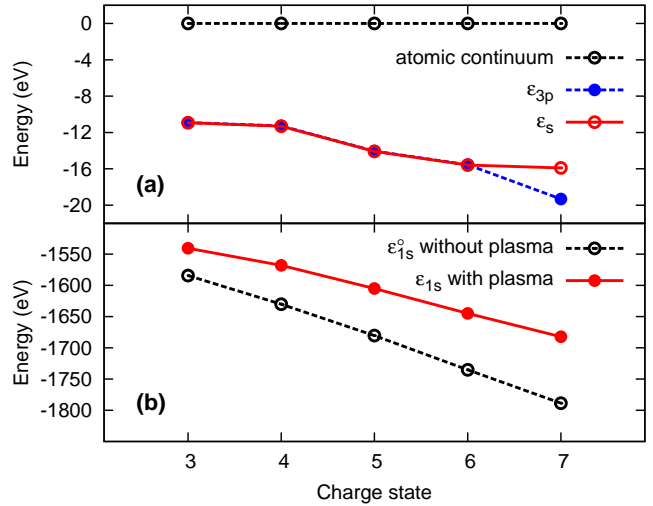}
\caption{\label{fig:Al_energy}%
Energy levels as a function of the charge state for aluminum.
(a) The $3p$ orbital energy and the threshold energy to the continuum ($\varepsilon_s$) calculated with the two-step HFS approach. 
The threshold energy to the continuum for isolated ions ($\varepsilon=0$) is shown for reference.
(b) The $1s$ orbital energies without the plasma (unscreened HFS calculations) and with the plasma (two-step HFS calculations).
}
\end{figure}

Figure~\ref{fig:Al_energy} shows (a) the resulting $3p$ (or the lowest-energy $p$ state in the continuum) orbital energies and the muffin-tin flat potential calculated by the two-step HFS scheme with the free-electron density, and (b) the $1s$ orbital energies with and without the plasma, as a function of the charge state. All those energies are lowered as the charge state increases. 
Note that the $3p$ energy lies right at the threshold to the continuum, i.e., it is not bound to a single atom. 
The only exception is Al$^{7+}$, where $3p$ lies $\sim$3.4~eV below the threshold. For an isolated atom or ion, calculated in the unscreened HFS approach, the threshold energy to the continuum is constant ($\varepsilon=0$) for all charge states. For a solid, the threshold energy to the continuum ($\varepsilon_s$) decreases by 5~eV from Al \textsc{iv} to Al \textsc{viii}. 
Lowering of the $1s$ binding energies due to the plasma environment (44--107~eV) is much larger than the lowering of the threshold energy (11--16~eV).
For $T>0$~eV, the difference between the threshold energy to the continuum in Fig.~\ref{fig:Al_energy}(a) and the $1s$ orbital energy in Fig.~\ref{fig:Al_energy}(b) gives the $K$-shell ionization potential.
In our approach, both the $1s$ orbital energy and the threshold energy are modified by the plasma environment. 

\begin{figure}
\centering
\includegraphics[width=\figurewidth]{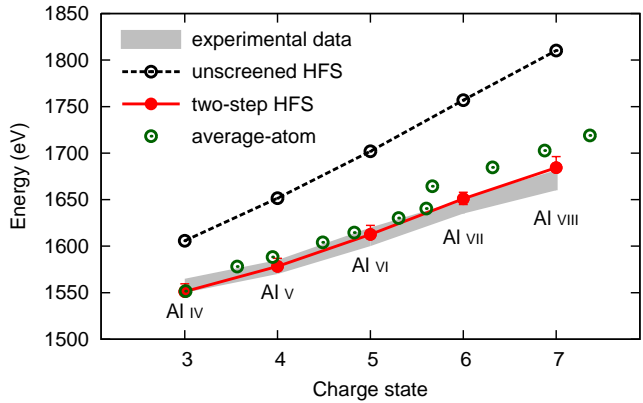}
\caption{\label{fig:fixed_config}%
$K$-shell thresholds for aluminum as a function of the charge state calculated with the two-step HFS model. 
The experimental data are taken from Ref.~\cite{Ciricosta12}. 
Unscreened HFS refers to calculations for isolated ions.}
\end{figure}


In the LCLS experiment on Al plasma~\cite{Ciricosta12}, $K\alpha$ fluorescence was detected and spectrally resolved as a function of the incoming photon energy. 
In this way, the incident-photon-energy threshold for the formation of a $K$-shell hole was determined for each energetically resolvable charge state. 
A $K$-shell hole can be created for $T>0$~eV by inner-ionization or photo-excitation into the $3p$ orbital if $3p$ is bound. 
%
Figure~\ref{fig:fixed_config} shows the calculated $K$-shell ionization thresholds (photo-excitation for Al$^{7+}$), in comparison with the experimental results~\cite{Ciricosta12}. 
We also plot the $K$-shell ionization thresholds for the unscreened HFS method (isolated ions) and the average-atom model.
For Al$^{7+}$, the resonant excitation into $3p$ is below the ionization threshold by just $\sim$3.4~eV, which may not be resolvable due to the LCLS energy bandwidth of $\sim$7~eV in experiment~\cite{Vinko12}.
As shown in Fig.~\ref{fig:fixed_config}, the two-step HFS calculation yields good agreement with the experimental data. 
However, the average-atom model alone fails in reproducing experiment, especially for high charge states.
In experiment, each discrete fluorescence line selects only one charge state and the $K$-shell threshold is assigned to this specific charge state.
The fixed-configuration scheme in our two-step model properly describes this selection of the $K$-shell threshold, whereas the average-atom model with the configuration averaging does not.
All calculated energies were shifted by +21.5~eV, according to the difference between the inner-ionization energy calculated at $T=0$~eV (1538.1~eV) and the experimental binding energy (1559.6~eV)~\cite{Thompson01}.
This constant energy shift is a model assumption for comparing our results to the experimental data.
Note that the absolute accuracy of HFS binding energies is typically about 1\%.
Clearly, in order to improve the description, one would require a treatment of the electronic structure beyond the mean-field level. 
However, it may be anticipated that such an approach would be much less efficient than the present HFS theory.
%
On the other hand, the error bar in the two-step HFS model in Fig.~\ref{fig:fixed_config} indicates variation from different thermal exchange potentials used in our calculations.
We have tested the thermal exchange potentials of \citet{Perrot84} and \citet{Rozsnyai72}, following the same two-step procedure as described in Sec.~\ref{sec:theory}, and found that the maximum deviation from the results with the zero-temperature potential is about 12~eV.


\begin{figure}
\centering
\includegraphics[width=\figurewidth]{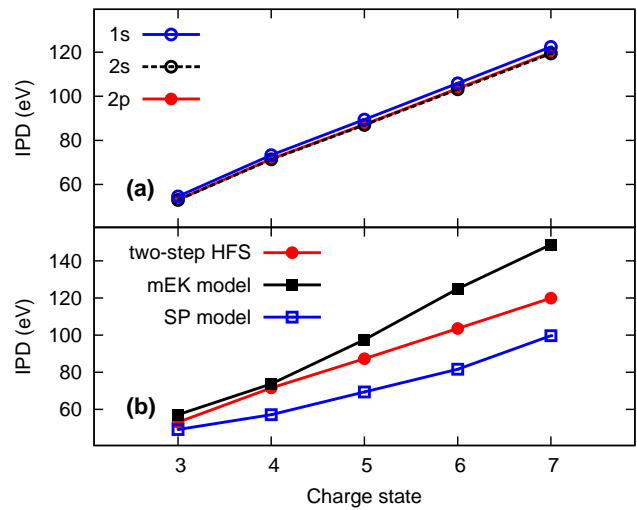}
\caption{\label{fig:IPD}%
Plots of the ionization potential depression.
(a) Comparison for all orbitals of Al calculated with the two-step HFS approach.
(b) Comparison of various methods: two-step HFS model, modified EK model, and SP model. The predictions of the latter two are taken from Ref.~\cite{Ciricosta12}.
}
\end{figure}

By taking the difference of the ionization potentials with and without the plasma environment, we can examine the lowering of the ionization potentials not only for $K$-shell electrons but also for electrons in other subshells. 
For individual bound orbitals of Al ions, we obtain the IPD shown in Fig.~\ref{fig:IPD}(a).
For isolated atoms, the ionization potential is given by $-\varepsilon^\circ_j$ where $\varepsilon^\circ_j$ is the $j$th orbital energy from the unscreened HFS calculation. For atoms in the plasma for $T>0$~eV, the inner-ionization potential is calculated by $\varepsilon_s - \varepsilon_j$, where both $\varepsilon_s$ and $\varepsilon_j$ are obtained from the two-step HFS calculation. Since the plasma screening affects each orbital differently, it is expected that IPDs for individual orbitals are different. Our results show that the IPD for $1s$ is higher than the IPD of $2s$ and $2p$ by $\sim$3~eV, but there is almost no difference in the IPDs for orbitals with the same principal quantum number ($2s$ and $2p$). This trend is similar to that observed in Ref.~\cite{Thiele12} for the Debye-screened HFS model.

Figure~\ref{fig:IPD}(b) depicts a comparison of various theoretical IPD models.
The results of the Stewart-Pyatt (SP) model and the modified Ecker-Kr\"oll (EK) model are taken from Ref.~\cite{Ciricosta12}. 
The original EK model~\cite{Ecker63} and the SP model~\cite{Stewart66} for lowering of the ionization energy have been widely used in the past decades and are implemented in several codes, e.g., \textsc{flychk}~\cite{Chung05} or \textsc{lasnex-dca}~\cite{Lee87}.
Ecker and Kr\"oll~\cite{Ecker63} have described lowering of the ionization potential as being due to the presence of an electric microfield. 
In their model, there is no difference among the IPDs of individual orbitals, and the ionization potential is considered as the difference between the ground-state energy of the charge state $Q$ and that of the charge state $(Q+1)$, which corresponds to the outermost-shell ionization potential in our calculations ($2p$ for the Al plasma). 
A modified version of the EK model (mEK) has been proposed in Refs.~\cite{Ciricosta12,Preston13} by employing an empirical constant to fit the experimental data. 
Figure~\ref{fig:IPD}(b) shows that neither the mEK model nor the SP model are close to our two-step HFS approach.

\begin{figure}
\centering
\includegraphics[width=\figurewidth]{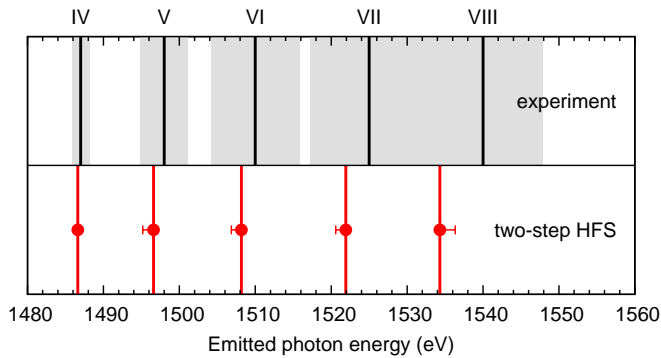}
\caption{\label{fig:transition_energy}%
$K\alpha$ transition energies of aluminum ions obtained with the two-step HFS model. 
They are compared to the experimental data taken from Ref.~\cite{Ciricosta12}.
The gray areas indicate the full-width-at-half-maximum from experiment.
}
\end{figure}

In Figure~\ref{fig:transition_energy}, we show the peak positions of the $K\alpha$ fluorescence lines for Al \textsc{iv} up to Al \textsc{viii}.  
The experimental data are taken from Ref.~\cite{Ciricosta12}. 
The transition energies are calculated from the differences of the $2p$ and $1s$ orbital energies in the fixed-configuration scheme. 
The fixed bound-electron configuration associated with a given charge state is chosen from Table~\ref{table:config}.
All calculated energies are shifted by +21.5~eV, according to the difference between the Al \textsc{iv} transition energy calculated with the average-atom model at $T=0$~eV ($1464.9$~eV) and the experimental transition line ($1486.4$~eV)~\cite{Thompson01}. 
The error bar in the two-step HFS model shows variation from usage of different thermal exchange potentials, and the maximum deviation is about 2~eV.
Our results show small deviations ($<5.7$~eV) from the experimental transition energies.
Here we show transition energies for only one configuration associated with a given charge state.
However, different configurations, as listed in Table~\ref{table:config_probability} at $T=80$~eV, would give rise to different transition lines.
For instance, the ground configuration ($1s^2 2s^2 2p^2$) of Al~\textsc{viii} at $T=80$~eV gives a fluorescence line of +3~eV higher in energy than the most probable configuration ($1s^2 2s^1 2p^3$).

\subsection{Al plasmas at high temperature and high density: Orion experiment}

\begin{figure}
\centering
\includegraphics[width=\figurewidth]{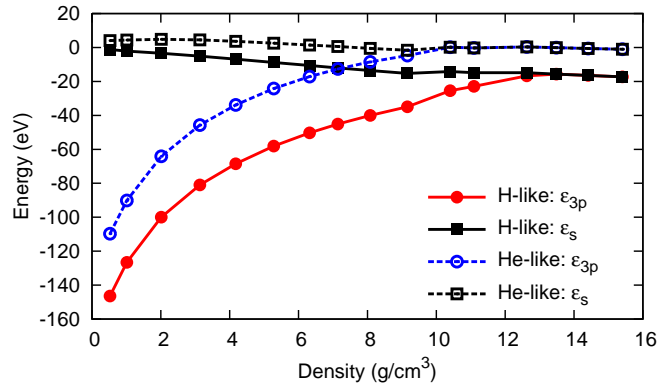}
\caption{\label{fig:orion}%
The muffin-tin flat potential ($\varepsilon_s$) and $3p$ orbital energy ($\varepsilon_{3p}$) of H-like and He-like Al as a function of the solid density at $T=700$~eV.}
\end{figure}

Our two-step scheme is applicable not only in the strongly coupled plasma regime but also in the weakly coupled plasma regime.
Recently, \citet{Hoarty13,Hoarty13a} used the high-power Orion laser to create compressed Al plasmas with high temperature and measured Ly$_\beta$ and He$_\beta$ lines to diagnose the Al plasmas created.
In Fig.~\ref{fig:orion}, the two-step model calculations show that the $3p$ state of compressed Al merges with the continuum ($\varepsilon \geq \varepsilon_s$) as the Al density increases.
The electronic temperature is 700~eV, close to the Orion experimental condition~\cite{Hoarty13}.
For the bound-electron part in the second step, the H-like Al is calculated with the exact one-electron potential and the He-like Al is calculated with the exact Hartree-Fock potential.
For such a high temperature, the plasma electron density contributes to only the direct Coulomb interaction~\cite{Rozsnyai72}.

When the solid density $d$ is greater than $\sim$12~g/cm$^3$, the $3p$ state of H-like Al is no longer bound, so the Ly$_\beta$ line would disappear.
Likewise, the He$_\beta$ line would disappear after $d>10$~g/cm$^3$. 
Our results are consistent with the experimental finding of no $n=3$ transitions occurring at $d>8$--10~g/cm$^3$.
When we use different thermal exchange potentials~\cite{Perrot84,Rozsnyai72}, the merging point of $3p$ becomes $\sim$8~g/cm$^3$ and $\sim$11~g/cm$^3$ for He$_\beta$ and Ly$_\beta$, respectively.
The SP model predicts delocalization of $n=3$ levels for $d>11.6$~g/cm$^3$, whereas the EK model prediction was found to be in clear disagreement with the experimental data~\cite{Hoarty13}.

\section{Conclusion}\label{sec:conclusion}
In this work, we have extended the standard HFS approach for calculating atomic energy levels for ions embedded in a plasma, taking into account plasma screening. 
Our two-step HFS model includes: (i) the average-atom model to obtain the free-electron density at a given temperature and (ii) the fixed-configuration model taking into account the free-electron density. 
Our current analysis focused on Al plasmas created by LCLS~\cite{Vinko12,Ciricosta12} and Orion laser~\cite{Hoarty13}, covering both strongly and weakly coupled plasma regimes.
Our two-step HFS results on the $K$-shell threshold energies of different charge states within Al plasma are in good agreement with the LCLS experimental data~\cite{Vinko12,Ciricosta12}.
References~\cite{Vinko12,Ciricosta12} measured fluorescence threshold energies that were then used to extract IPDs by combining the data with a specific theory model describing the unscreened ionization potentials. 
Thus, the estimated IPDs relied on the theory of the unscreened case, which hinders direct comparison with IPD models.
In contrast, our model computes the energy shifts of all individual orbitals with and without plasma screening, thus providing IPDs in an internally consistent manner. 
Our calculated valence IPDs lie between the SP and mEK models. 
Hence, we cannot confirm that the performance of the mEK model is superior to that of the SP model, as suggested in Ref.~\cite{Ciricosta12}.  
Moreover, in the high temperature regime, our prediction for the $3p$ state is in good agreement with the SP model and reproduces Orion experimental data~\cite{Hoarty13}.
These results show that, with our proposed two-step HFS approach, a reliable and relatively inexpensive calculation of atomic properties within plasmas can be performed both for weakly and strongly coupled plasmas. 
We therefore expect that our model will be a useful tool for describing new data from plasma spectroscopy experiments.


\begin{acknowledgments}
We thank Hyun-Kyung Chung, Orlando Ciricosta, Nikita Medvedev, Sam M.\ Vinko, and Justin S.\ Wark for helpful discussions.
During the review process, a new preprint treating the same IPD problem appeared (B.\ J.\ B.\ Crowley, arXiv:1309.1456).
\end{acknowledgments}

\appendix
\section{Probability distributions of charge states and configurations}\label{appendix1}
The partition function of the grand-canonical ensemble for fermions is given by
\begin{align}
Y &= \mathrm{Tr}\{e^{-\beta(\hat{H}-\mu\hat{N})}\}
\nonumber
\\
&= \sum_{ \{ n_p \} }^\text{all} e^{-\beta\sum_p(\varepsilon_p-\mu)n_p} 
\nonumber
\\
&= \prod_{p=1}^{\infty} \left( 1 + e^{-\beta(\varepsilon_p-\mu)} \right),
\end{align}
where $\beta=1/T$ is the inverse of the temperature, $\{n_p\} = ( n_1, n_2, \cdots )$, and $n_p$ is the occupation number of the $p$th one-electron state.
Here $p$ includes all bound and continuum one-electron states, and the summation runs over all possible configurations $\{ n_p \}$.
For fermions, $n_p$ is either 0 or 1.
%
Then, the probability of finding one specific configuration $\{n_p\}$ is
\begin{align}\label{eq:P_one_config}
P_{\{n_p\}} = \frac{1}{Y} e^{-\beta\sum_p (\varepsilon_p - \mu) n_p} = \prod_{p=1}^{\infty} \frac{ e^{-\beta(\varepsilon_p-\mu) n_p}}{ 1 + e^{-\beta(\varepsilon_p-\mu)} }.
\end{align}
%

Let us consider a fixed configuration, more precisely, a fixed bound-electron configuration $[n_b] = ( n_1, \cdots, n_B )$, where $B$ is the number of the bound states.
A general Fock-space configuration consistent with the fixed bound-electron configuration is  
\begin{equation}
\{ n_b; n_{p'} \} = ( n_1, \cdots, n_B; n_{B+1}, \cdots ).
\end{equation}
Here $n_b$ for $1 \leq b \leq B$ is a fixed occupation number, while $n_{p'}$ is either 0 or 1 for $p' \geq B+1$.
The probability of finding the fixed configuration $[n_b]$ is calculated by summing over all such configurations $\{ n_b; n_{p'} \}$,
\begin{align}
P_{[n_b]} 
&= \sum_{\{ n_p \} = \{ n_b; n_{p'} \} } P_{\lbrace n_p \rbrace}
\nonumber
\\
&= 
\frac{1}{Y} \prod_{b=1}^B e^{-\beta(\varepsilon_b - \mu) n_b} \sum_{ \{ n_{p'} \} }^\text{all} \prod_{p'=B+1}^{\infty} e^{-\beta(\varepsilon_{p'} - \mu) n_{p'}}
\nonumber
\\
&= 
\frac{ 
\prod_{b=1}^B e^{-\beta(\varepsilon_b - \mu) n_b} \prod_{p'=B+1}^{\infty} \left( 1 + e^{-\beta(\varepsilon_{p'} - \mu)} \right)
}{ 
\prod_{b=1}^B \left( 1 + e^{-\beta(\varepsilon_b - \mu)} \right) \prod_{p'=B+1}^{\infty} \left( 1 + e^{-\beta(\varepsilon_{p'} - \mu)} \right)
}.
\nonumber
\end{align}
In the above calculation, since the occupation numbers in the bound states are fixed, they can be factored out and the remaining parts in the numerator and the denominator are canceled out.
Thus, the probability reduces to
\begin{align}
P_{[n_b]} 
&= 
\prod_{b=1}^{B}\frac{ e^{-\beta(\varepsilon_b-\mu) n_b}}{1 + e^{-\beta(\varepsilon_b-\mu)}},
\end{align}
which corresponds to Eq.~\eqref{eq:prob_bound_config}.
This is for spin-orbitals $b$, i.e., $n_b$ is either 0 or 1.
For practical purposes, we express the probability in terms of subshells,
\begin{equation}
P_{[n_i]} = \frac{ 
\prod_{i=1}^{N_b} \binom{ 4 l_i + 2 }{ n_i } e^{ -\beta ( \varepsilon_{i} - \mu ) n_i }
}{ 
\prod_{i=1}^{N_b} \left( 1 + e^{ -\beta ( \varepsilon_{i} - \mu ) } \right)^{(4 l_i + 2)}
},
\end{equation}
where $i$ is the subshell index, $N_b$ is the number of bound subshells, and $[n_i] = ( n_1, \cdots, n_{N_b} )$.
For the $i$th subshell, $\varepsilon_i$ is the orbital energy, $l_i$ is the orbital angular momentum quantum number, and $n_i$ is the occupation number ($0 \leq n_i \leq 4 l_i + 2$).

The probability of finding the charge state $Q$ is given by summing over all possible configurations associated with $Q$,
\begin{equation}\label{eq:P_Q}
P_Q = \sum_{[n_b]}^Q P_{[n_b]} = \sum_{[n_b]}^Q \prod_{b=1}^{B} \frac{ e^{-\beta(\varepsilon_b-\mu) n_b}}{1 + e^{-\beta(\varepsilon_b-\mu)}},
\end{equation}
%
where $[n_b]$ satisfies $\sum_{b=1}^{B} n_b = Z - Q$.  
This is Eq.~\eqref{eq:prob_charge}.
It is straightforward to verify $\sum_Q P_Q = 1$. 
The average charge state $\bar{Q}$ is given by
\begin{align}
\bar{Q} 
&= \sum_Q Q P_Q 
\nonumber
\\
&= \sum_Q \sum_{[n_b]}^Q \left( Z - \sum_{p=1}^{B} n_p \right) \prod_{b=1}^{B} \frac{ e^{-\beta(\varepsilon_b-\mu) n_b} }{ 1 + e^{-\beta(\varepsilon_b-\mu)} }
\nonumber
\\
&= Z - \sum_{b=1}^B \tilde{n}_b(\mu,T),
\end{align}
which is used to calculate $\bar{Q}$ in Tables~\ref{table:Al} and \ref{table:config}.

\section{Determination of free-electron density}\label{appendix2}
Here we calculate the total electron density for a fixed bound-electron configuration, $[ n_b ] = ( n_1, \cdots, n_B )$.
The total density is chosen by averaging densities over all configurations $\{ n_b; n_{p'} \}$ that have $[ n_b ]$ in common:
\begin{equation}\label{eq_A2:rho_nb}
\rho_{[n_b]} = \sum_{ \{ n_p \} = \{ n_b; n_{p'} \} } \rho_{\{ n_p \} } w_{ \{ n_p \} },
\end{equation}
where $w_{ \{ n_p \} }$ is a statistical weight,
\begin{align}\label{eq_A2:w}
w_{ \{ n_b; n_{p'} \} } 
&= \frac{ e^{-\beta\sum_p (\varepsilon_p - \mu) n_p} }{ \sum_{ \{ n_p \} = \{ n_b; n_{p'} \} } e^{-\beta\sum_p (\varepsilon_p - \mu) n_p} }
\nonumber
\\
&= \prod_{p'=B+1}^{\infty} \frac{ e^{- \beta ( \varepsilon_{p'} - \mu ) n_{p'} } }{ 1 + e^{- \beta ( \varepsilon_{p'} - \mu ) } }.
\end{align}
The total density for $\{ n_p \}$ decomposes into bound-electron and free-electron densities,
\begin{align}\label{eq_A2:rho_np}
\rho_{\{ n_p \}} &= \sum_{b=1}^B \left| \psi_b(\mathbf{r}) \right|^2 n_b + \sum_{p'=B+1}^{\infty} \left| \psi_{p'}(\mathbf{r}) \right|^2 n_{p'}.
\end{align}
Plugging Eqs.~\eqref{eq_A2:w} and \eqref{eq_A2:rho_np} into Eq.~\eqref{eq_A2:rho_nb}, we evaluate the total density for $[n_b]$.
The bound-electron density for $[n_b]$ is then
%
\begin{align}
\rho_{b,[n_b]} &= \left( \sum_{b=1}^B \left| \psi_b(\mathbf{r}) \right|^2 n_b \right) \sum_{ \{ n_{p'} \} }^\text{all} \prod_{p'=B+1}^{\infty} \frac{ e^{- \beta ( \varepsilon_{p'} - \mu ) n_{p'} } }{ 1 + e^{- \beta ( \varepsilon_{p'} - \mu ) } }
\nonumber
\\
&= \sum_{b=1}^B \left| \psi_b(\mathbf{r}) \right|^2 n_b,
\end{align}
and the free-electron density for $[n_b]$ is given by
\begin{align}
\rho_{f,[n_b]} &= \sum_{ \{ n_{p'} \} }^\text{all} \!\! \left( \sum_{p'=B+1}^{\infty} \left| \psi_{p'}(\mathbf{r}) \right|^2 n_{p'} \!\! \right) \!\! \prod_{p'=B+1}^{\infty} \frac{ e^{- \beta ( \varepsilon_{p'} - \mu ) n_{p'} } }{ 1 + e^{- \beta ( \varepsilon_{p'} - \mu ) } }
\nonumber
\\
&= \sum_{p=B+1}^{\infty} \left| \psi_{p}(\mathbf{r}) \right|^2 \frac{ 1 }{ e^{\beta ( \varepsilon_{p} - \mu ) } + 1}.
\end{align}
Therefore, the total electron density for configuration $[n_b]$ is determined from the average-atom calculation (the first step of our two-step approach) with the grand-canonical ensemble,
\begin{align}
\rho_{[n_b]} &= \rho_{b,[n_b]} + \rho_{f,[n_b]}
\\
&= \sum_{b=1}^B \left| \psi_b(\mathbf{r}) \right|^2 n_b + \sum_{p=B+1}^{\infty} \left| \psi_{p}(\mathbf{r}) \right|^2 \tilde{n}_p(\mu,T).
\end{align}
In the second step of our two-step approach, we separate out the free-electron density, $\rho_{f,[n_b]}$, which yields Eq.~\eqref{eq:rho_f}.


%


\end{document}